\documentclass[aps, pre, floatfixt,12ptsi]{revtex4-1}
\usepackage[pdftex]{graphicx}
\usepackage{amsmath}
\usepackage{amssymb}
\usepackage{tabulary}
\begin{document}

\title[Thermomagnetic stability in type-II superconductors]{Thermomagnetic stability in isotropic type-II superconductors under multicomponent
magnetic fields}
\author{O.A. Hern\'andez-Flores, J.
Guti\'errez-Guti\'errez and C. Romero-Salazar}
\address{Ingenier\'ia en Innovaci\'on Tecnol\'ogica, 
Universidad Aut\'onoma Benito Ju\'arez de Oaxaca, Av. Universidad s/n Col. Cinco Se\~nores. C.P. 68120. Oaxaca, Oax. M\'exico }
\email{ohernandez.ciencias@uabjo.mx}
\begin{abstract}
The goal of this research is the study of the thermomagnetic consequences in isotropic type-II superconductors, subjected to multi-component magnetic fields $\mathbf{H}_a = H_{ay}\hat{y}+H_{az}\hat{z}$, because the instability field $\mathbf{H}_{fi}$ is closely related with a flux jump occurrence. At the critical-state model framework, once the Lorentz $\mathbf{F}_L$ and pinning forces $\mathbf{F}_P$ are at equilibrium, the current density reaches a critical value $j_c$ and a stationary magnetic induction distribution $\mathbf{B}$ is established. The equilibrium of forces is analytically solved considering that the pinning force is mainly affected by temperature increments; the energy dissipation is incorporated throughout the heat equation at the adiabatic regime. The theory is able to obtain the instability field according to the thermal bath and applied field values; moreover, it provides of instability field branches comprising both partial and full penetrates states. With this information is possible to construct a field-temperature map. The results are compared with already published experimental data, finding a qualitatively agreement between them. This theoretical study works with a first order perturbation, then the perturbation presents a periodical behavior along the thickness direction; considering this environment it is constructed the magnetic induction distributions which resemble flexible cantilever structures.
\end{abstract}


\maketitle
%
\section{Introduction}\label{intro}
%
Since a half century to find criteria to realize and control how type-II superconductors remain in
a stable state, with the maximum trapped flux and the minimum of losses, has
been a research subject. For each value of the magnetic induction in the range
$B_{c1} < B < B_{c2}$ there is a specific maximum value of the current density $j_c$ which if it
is exceeded causes that the superconductor enters into a resistive state up to change its phase and 
enters into the normal state. The current density $j_c(B,T)$ is called the critical value for specific
values of the induction B and temperature T. 

At present, it is clear the Lorentz force $F_L$ nature as well as it drives the
vortices to move across the sample, however, the pinning force $F_P$ origin is
so complex that an exhaustive description requires another studies, out of those
presented here; for the time being it is enough to model $F_P$ as an average of an uniform
distribution of pinning sites, qualitatively described by the
Kim-Anderson model for $j_c(B,T)$ \cite{PhysRev.129.528,PhysRevB.42.10773}.

There are a list of facts related with the type-II superconductors behavior: at
the equilibrium of forces, the magnetic induction is smooth and the slope at
each point of the sample corresponds to a critical current density $j_c$, which
generally speaking depends on the magnetic induction and has a curvature; on
increasing the current density above a certain maximum value the Lorentz force
becomes greater than the pinning force, so the vortex starts to move; this leads
to the continuous evolution of Joule's heat, the temperature may rise above $T_c$,
and the superconductor will enter into the normal state; the material stability
depends on the ramp rate of the applied magnetic field, the temperature, the
specif heat of the material, the heat release conditions of the experimental set
up, the geometry as well as the preparation sample. Because such variables are
correlated it is desirable to know its participation on such process \cite{al2013stabilization}.

In the present article, the balance of forces scheme is used to describe the problem of instabilities \cite{sosnowski1983flux, sosnowski:jpa-00245326, sosnowski1986influence,PhysRev.161.404, wipf1991review}.
It is knows that the thermomagnetic stability description requires to take into account the equilibrium between the Lorentz, pinning and viscous forces.
The external field drives the vortices throughout the Lorentz force
$\mathbf{F}_L=\mathbf{j} \times \mathbf{B}=(1/\mu_0)\nabla \times \mathbf{B}
\times \mathbf{B}$ and their movement can be prevented by pinning $\mathbf{F}_P$
or viscous $\mathbf{F}_{\mathit{v}}$ forces, therefore the critical state corresponds to the
equilibrium forces $\mathbf{F}_L+\mathbf{F}_P+\mathbf{F}_v=\mathbf{0}$. 
The viscous force is a dynamical force which depends on the vortex velocity, in our theory
we assume that the vortex velocity is such that the viscous force is not comparable with the
pinning force therefore its contribution is neglected.  

Therefore, we study an isotropic superconductor embedded in a thermal bath
with temperature $T_B$ and subject to an external magnetic field $H_a$, 
for each external magnetic field change, the flux front
goes deep forward inside up to a $x_0$ point. 
As $H_a$ changes as $H_a+\Delta H_a$ the temperature changes as $T_B+\Delta T$
due to Joule's heat, since the magnetic diffusion time is much smaller than the 
thermal diffusion time, the magnetic induction reaches an stationary state before the temperature
be homogenized at the thermal bath. A consequence of this fact is that it decreases the critical current density, promoting that the flux front goes deeper across the sample.

The equilibrium force condition provides of isothermic profiles of magnetic induction
$B_{i}(x)$, in our study we are going to consider that an increment $\Delta H_a$
will produce a new magnetic profile, given by $\mathbf{B}=\mathbf{B}_i + \Delta \mathbf{B}$.
At this new configuration, the Lorentz and pinning forces are written as 
$\mathbf{F}_L(\mathbf{B}_i + \Delta \mathbf{B}) = \mathbf{F}_L(\mathbf{B}_i) +
\Delta \mathbf{F}_L(\mathbf{B}_i, \Delta \mathbf{B})$ and
$\mathbf{F}_P(\mathbf{B}_i + \Delta \mathbf{B}) = \mathbf{F}_P(\mathbf{B}_i) +
\Delta \mathbf{F}_P(\mathbf{B}_i, \Delta \mathbf{B})$, where the condition
$\Delta \mathbf{F}_L+ \Delta \mathbf{F}_P=\mathbf{0}$ is fulfilled.

The changes on the external field produce thermal variations which can produce instabilities on the system,
however, the magnetic properties of the superconducting material will stay stable under 
such changes meanwhile the pinning force subjects the Lorentz force, under these conditions the net transport of vortices is null. The stability condition can be written as follows:
\begin{eqnarray}
|\mathbf{F}_L(\mathbf{B}_i) + \Delta \mathbf{F}_L(\mathbf{B}_i, \Delta
\mathbf{B})|
\underset{\mathrm{Instability}}{\overset{\mathrm{Stability}}{\lessgtr}} 
|\mathbf{F}_P(\mathbf{B}_i) + \Delta \mathbf{F}_P(\mathbf{B}_i, \Delta
\mathbf{B})|
\label{e0}
\end{eqnarray}

The aim of this article is to find the instability magnetic field
$H_{fi}$, which delimits the magnetic stability of the superconductor, defined by the
condition (\ref{e0}).
It is considered a semi-infinite-isotropic type-II superconductor plate, with
thickness $d$ along the $x-\mathrm{direction}$ and its surface lying on the $yz-\mathrm{plane}$, under the effect of an applied magnetic field $\mathbf{H}_a = H_{ay}\hat{y} + H_{az}\hat{z}$. At this situation, the so-called parallel geometry, all the electromagnetic vectors  are coplanar to the $yz$ surface and functions only of the variable $x$. It is assumed that the applied field is larger than the first critical field $H_{c1}$, thus the approximation $\mu_0\mathbf{H}a=\mathbf{B}(x=0)=\mathbf{B}(x=d)$ is fullfiled.

This manuscript is divided as follows: In section (\ref{sec:2}) one can find an 
analytic expression for the isothermal
magnetic induction $\mathbf{B}_i$ obtained from the balance of forces; in order to
simplify the notation, it is removed the subindex $``i"$ since the main purpose of this paper is 
the resulting field $\mathbf{B}_i+\Delta\mathbf{B}$. In section (\ref{sec:3}) throughout the
deviation forces balance, after a small increment of the applied magnetic field
occurs, is obtained expressions for $\Delta B$ and 
$\mu_0H_{fi}$ at the adiabatic regime. Finally, in section (\ref{sec:4}) it is presented
 an example where it is compared the theory proposed and experimental data, obtaining an $H-T$ instability diagram, $B$ and temperature profiles.

%
\section{Equilibrium between Lorentz and pinning forces and the isothermal
approach}\label{sec:2}
%

This teoretical development starts considering the equilibrium between Lorentz and
pinning forces $\mathbf{F}_L=-\mathbf{F}_P$; the external field drives the vortices motion through the Lorentz force $\mathbf{F}_L=\mathbf{j} \times \mathbf{B}=(1/\mu_0)\nabla
\times \mathbf{B} \times \mathbf{B}$, however, such movement can be prevented by
pinning centers represented with an average force knows as the pinning force  $\mathbf{F}_P$. 
In the following lines, is it shown explicit expressions for both forces in terms of the magnetic
induction. From the balance of forces it will obtain the equation which governs the magnetic induction. Let us show this fact.

Writing the magnetic induction  as $\mathbf{B}=B(x) \hat{e}$, with $
\hat{e}=\cos\phi\hat{x}+\sin\phi\hat{y}$, since the system is at the parallel geometry,
the Ampere's law looks like:

\begin{eqnarray}
\mu_0 \mathbf{j}=B \left(-\frac{d \phi}{dx}\right) \hat{e} - \frac{dB}{dx}
(\hat{e} \times \hat{x})
\label{e2-1}
\end{eqnarray} 
The Lorentz force is 
\begin{eqnarray}
\mu_0 \mathbf{F}_L=-B^2 \frac{d\phi}{dx} \hat{e} \times
\hat{e}-B\frac{dB}{dx} (\hat{e} \times \hat{x}\times \hat{e})=-B \frac{dB}{dx}
\hat{x},
\label{e2-2}
\end{eqnarray}
while the pinning force is modelled as
\begin{eqnarray}
\mu_0\mathbf{F}_P=\mu_0 j B \cos\theta \hat{x};
\label{e2-3}
\end{eqnarray}
here $\theta$ is the angle of $\mathbf{j}$ respect to $\mathbf{B}$.  Notice the change of sign of the equation (\ref{e2-3}) according to
\begin{eqnarray}
\cos \theta = \left\{ \begin{array}{lr}
  <0,\quad & 0\leq x <d/2 \\
  >0,\quad & d/2<x\leq d 
     \end{array}
      \right.,       
\end{eqnarray}
 therefore when the sample is at critical state $j=j_c$ and $\theta=0,\pi$ at $0\leq x <d/2$, $d/2<x\leq d$, respectively.
It is defined the function $f_H=2H_e(x-d/2)-1$, where $H_e(x-d/2)$ is the
Heaviside function, to rewrite the equation (\ref{e2-3}) as $\mu_0 \mathbf{F}_P
=\mu_0j_c Bf_H\hat{x}$. It is employed the well-known phenomenological relation for
the critical current density $j_c(B,T)$:
\begin{eqnarray}
j_c(B,T)=\alpha(T) \frac{j_0}{\left( 1+\frac{B}{B^*} \right)^n}
\end{eqnarray}
where $\alpha$ is a function of the temperature; $n$
and $B^*$ are parameters.
Now, defining $b=1+B/B^*$, the pinning and Lorentz forces acquire the form
\begin{eqnarray}
\mu_0\mathbf{F}_P=\alpha(T)\mu_0j_0B^* \frac{(b-1)}{b^n}f_H \hat{x},
\quad
\mu_0\mathbf{F}_L=-(b-1)\frac{db}{dx} (B^*)^2 \hat{x},
\label{e2-5}
\end{eqnarray}
with the forces balance, it is found the equation
that governs the magnetic induction behavior
\begin{eqnarray}
\frac{db}{dx}=\frac{\alpha(T)\mu_0j_0}{B^*}\frac{f_H}{b^n}.
\label{e2-6}
\end{eqnarray}
For the research purposes of this work, in the following lines, isothermic magnetic induction profiles will be obtained; in this case $\alpha$ is constant.

 The ordinary differential equation (\ref{e2-6}) is solved considering the boundary conditions
$B(x=0)=B(x=d)=\mu_0H_a$. The analytic solution is:
\begin{eqnarray}
b^{n+1}=f_H\frac{\alpha\mu_0j_0(n+1)}{B^*} (x-x_H) 
+ b_a^{n+1},
\label{e2-7}
\end{eqnarray} 
here $b_a=1+\mu_0H_a/B^*$ and $x_H=d/2\,H_e(x-d/2)$. The field $b^{n+1}$
describes a master curve, which is a simple straight line
\cite{ROMERO-SALAZAR2013}. The sample symmetry makes that the flux front
($x_f$ for $x\in [0,d/2)$ and $x'_f=d-x_0$ for $x\in (d/2,d]$)
goes forward up to reach the middle as the applied field be equal to the penetration field; the flux front for field values larger than the penetration field is given by the constant value$x_f=x'_f=d/2$. One should notice that the solution
(\ref{e2-7}), valid for $x \in [0,d]$, includes non physical solutions for partial penetrate states if
$b^{n+1} <0$ . To avoid this issue, one should work
with heuristic solutions, valid for both states, as the following
\begin{eqnarray}
b^{n+1}  = \left\{ \begin{array}{lr}
1 												       				  &   x_f < x < d-x_f; \\
b_a^{n+1} -\displaystyle\frac{\alpha\mu_0j_0(n+1)}{B^*} x          &  0\leq x\leq x_f;  \\ 
b_a^{n+1}+\displaystyle\frac{\alpha\mu_0j_0(n+1)}{B^*} (x-\frac{d}{2}) &    d-x_f< x \leq d.
 \end{array}
      \right.      
\label{e2-11}
\end{eqnarray}

The magnetic induction components will not be necessary for the purposes of this work, however, they can be written as follows 
\begin{eqnarray}
b_y = b\cos\phi +(1-\cos\phi), \quad b_z = b\sin\phi +(1-\sin\phi) \nonumber
\end{eqnarray} 
where $\phi = -\tan^{-1}(H_{az}/H_{ay})$.

To finish this section, it is calculated the flux front and the penetration field.The first one
 $x_f=x_f(\mu_0H_a)$ can be obtained for both partial and full penetrate states.
For partial penetrate states one has that
$B(x=x_f)=0$ or $b(x=x_f)=1$, therefore, using the equation (\ref{e2-11}) is obtained that,
\begin{eqnarray}
x_f=(1-b_a^{n+1})\frac{B^*}{\alpha\mu_0j_0(n+1)}.
\label{e2-8}
\end{eqnarray}  
For full penetrate states the flux front is always at
$x_f=d/2$, then the field minimum $b_h=b(x_f=d/2)$ is
\begin{eqnarray}
b^{n+1}_h=b_a^{n+1}-\frac{\alpha\mu_0j_0(n+1)}{B^*}\frac{d}{2},
\label{e2-9}
\end{eqnarray}
where $b_h=1+B_h/B^*$ with $B_h=B(x=d/2)$. The first penetration field corresponds to
$b_h=1$,
\begin{eqnarray}
b_p^{n+1}=\alpha (n+1)\frac{B_0}{B^*}+1,
\label{e2-10}
\end{eqnarray}
where $B_0 = \mu_0j_0d/2$ is the  Bean first penetration field. 

%
\section{Fluctuations balance and the instability field $H_{fi}$} 
\label{sec:3}
%
%
\subsection{Fluctuations balance}
When a magnetic field $\mu_0 \mathbf{H}_a$ is applied on
a superconducting material, the Lorentz and pinning forces vary up to reach the equilibrium
$\mathbf{F}_L+ \mathbf{F}_P=0$. If $\mu_0 \mathbf{H}_a$ is increased  $\mu_0 \Delta\mathbf{H}_a$, the superconductor suffers a change of state related with Lorentz
and pinning forces changes $\Delta \mathbf{F}_L=\mathbf{F}^{'}_L-\mathbf{F}_L$ and $\Delta
\mathbf{F}_P=\mathbf{F}^{'}_P-\mathbf{F}_P$, here $\mathbf{F}'_L$ and
$\mathbf{F}'_P$ denote the Lorentz and pinning forces due to $\mu_0
\mathbf{H}_a+\mu_0\Delta \mathbf{H}_a$.  At the equilibrium of forces
$\mathbf{F}^{'}_L+ \mathbf{F}^{'}_P=0$ or $\Delta \mathbf{F}_L+\Delta \mathbf{F}_P=0$. The magnetic induction $\mathbf{B}$ corresponds to $\mu_0 \mathbf{H}_a$ while  $\mathbf{B}+\Delta\mathbf{B}$ it is the outcome of a perturbation $\mu_0 \Delta\mathbf{H}_a$ added to  $\mu_0\Delta\mathbf{H}_a$. Here $\Delta\mathbf{B}$ corresponds to the magnetic
induction deviation due to the perturbated material state, both have the same
direction $\mathbf{B}=B\hat{e}$ and $\Delta\mathbf{B}=\Delta B\hat{e}$,
consequently $\mathbf{B} + \Delta \mathbf{B} =(B+\Delta B)\hat{e}$.

Let us show expressions for both Lorentz and pinning forces changes. The 
Lorentz force change is obtained from $\mu_0\Delta \mathbf{F}_L =
\mu_0\mathbf{F}_L(\mathbf{B}+\Delta \mathbf{B})-
\mu_0\mathbf{F}_L(\mathbf{B})=\mu_0 \Delta F_L \hat{x}$, its unique component is
\begin{eqnarray}
\mu_0 \Delta F_L = -B\frac{d\Delta B}{dx} -\Delta B \frac{dB}{dx},
\end{eqnarray}
the quadratic term $\Delta B (d(\Delta B)/dx)$ will be neglected. Using the variable $b=1+B/B^*$ and $db/dx=\alpha \mu_0 j_0 f_H/B^*(1/b^n)$, the
former equation is rewritten as
\begin{eqnarray}
\mu_0 \Delta F_L=-\left\lbrace (b-1) \frac{d \Delta B }{db} +\Delta B
\right\rbrace \frac{\alpha(T)\mu_0j_0 f_H}{b^n}.
\label{e3-1}
\end{eqnarray}

It is assumed that the pinning force is a functional $F_P=F_P(T,B;x)$, then it has the total derivative of $F_P$ is $dF_p/dx=\partial_x F_P + \partial_T F_P dT/dx + \partial_B F_P dB/dx
$. Integrating it is obtained
\begin{eqnarray}
\mu_0\Delta F_p = \mu_0\!\!\!\!\!\!\!\!\int\limits_{F_P}^{F_P+\Delta
F_P}\!\!\!\!\!\!\!\!F'_PdF'_P \;\;=
\mu_0\!\!\!\!\!\int\limits_T^{T+\Delta T}\!\!\!\!\!\! \partial_{T '}F_P
dT'\;\; +
\mu_0\!\!\!\!\!\int\limits_B^{B+\Delta B}\!\!\!\!\!\! \partial_{B'}
F_P dB', \nonumber
\end{eqnarray}
where $\partial_xF_P=0$ because it does not depend explicitly on the position. It is
assumed that $\partial_{T}F_P$ and $\partial_{B}F_P$ stay approximately constants
during the integration. With these arguments an approximate expression of
the pinning force change is obtained:
\begin{eqnarray}
\mu_0 \Delta F_P 
& \approx & \partial_T(\mu_0 F_P) \Delta T + \partial_b (\mu_0 F_P) \frac{\Delta
B}{B^*} \nonumber \\
& =  & 
\mu_0 j_0 B^* f_H\partial_T\alpha \frac{b-1}{b^n} \Delta T  +  
\frac{\mu_0\alpha j_0 B^* f_H}{b^{n+1}} \left\lbrace (-n)(b-1) +b \right\rbrace
\frac{\Delta B}{B^*}
\label{e3-2}                 
\end{eqnarray}
The equilibrium between forces demands that $\mu_0 \Delta
\mathbf{F}_L=-\mu_0\Delta \mathbf{F}_P$, with $\Delta F_P=\Delta F_L$; using
(\ref{e3-1}) and (\ref{e3-2}) an ordinary differential equation is obtained for $\Delta B$:
\begin{eqnarray}
\frac{d}{db}(\Delta Bb^n) = \left(\frac{\partial_T\alpha}{\alpha}
B^*b^n\right)\Delta T.
\label{e3-3}
\end{eqnarray}
The ramping rate and magnetic induction history can be thermomagnetic instabilities
sources associated with the heat diffusion dynamics inside the material at the flux
creep regime \cite{Romanovskii201489}, as well as at the flux flow regime
\cite{Espinosa-Torres2015}. However, in our approach there is no ramping neither a trapped flux and the field $B + \Delta B$ is calculated at the adiabatic approach.

Because the temperature changes $\Delta T$ calculation requires special
attention (See the appendix). Since it is assumed the superconductor in an adiabatic regime,
$\Delta T$ is calculated with the equation
\begin{eqnarray}
\Delta T = \frac{1}{C} \Delta q=\frac{B^{*}}{\mu_0C b^n}
\int\limits_{b_f}^{b} \Delta Bb^n db .
\label{e3-4}
\end{eqnarray}
Substituting the latter equation into (\ref{e3-3}) it is found
\begin{eqnarray}
\frac{d(\Delta B b^n)}{db}=\left( \frac{\partial_T\alpha}{\alpha}\right)
\frac{(B^*)^2}{\mu_0 C} \int\limits_{b_f}^{b}\!\!\!\Delta B b'^n db'. \nonumber
\end{eqnarray}
One should take into account the physics of function $\alpha(T)$, it must be a decreasing function; at low temperatures promotes the superconductivity and close to the
threshold of the transition to a normal state reaches the zero value, suppressing the superconductor state.

To our knowledge the function $\alpha$ is monotone decreasing then the derivate is always negative and is fulfilled $\partial_T\alpha = -|\partial_T\alpha|$, this fact allows to write the field 
$B_T = \sqrt{\mu_0C\alpha/|\partial_T\alpha|}$ and 
\begin{eqnarray}
\frac{d(\Delta B b^n)}{db}=-\left(\frac{B^*}{B_T}\right)^2
\int\limits_{b_f}^{b}\!\!\!\Delta B b'^n db'.
\label{e3-5}
\end{eqnarray}
This equation defines the $\Delta B$ behavior, such deviation should obey the following conditions: 
i) is zero at the sample boundaries then $\mu_0H_a=B(x=0)=B(x=d)$; 
ii) at the flux front has a maximum value , that is $\Delta B(b(x_f))=D$. 
These conditions help to obtain the solution of equation (\ref{e3-5}).
%
\subsection{Solving for $\Delta B$}
%
In general, the heat capacity has thermal and magnetic dependence $C=C(T,H)$
\cite{Chaba2013}, however, the magnetic dependence will not be considered.

Assuming that $\Delta Bb^n$ is twice differentiable at $B\in
[B_h,\mu_0H_a]$, the first and second derivative of
$\Delta Bb^n$ are continous. Derivating the equation (\ref{e3-5}) with respect to
$b$ and due to the first integral theorem it is obtained
\begin{eqnarray}
\frac{d^2}{db^2}(\Delta B b^n)=-\left(\frac{B^*}{B_T}\right)^2 (\Delta B b^n)
\label{e3-5vis}
\end{eqnarray}
this result is the well known harmonic oscillator for 
$\Delta B b^n$. Under the hypothesis that $\Delta B b^n$ is twice
differentiable, $\Delta B b^n$ is continuous and therefore integrable. Such
arguments ensure that the problems (\ref{e3-5}) and (\ref{e3-5vis}) are equivalent. 
The general solution for the harmonic oscillator is
\begin{eqnarray}
\Delta B b^n = c_1\cos\left(\frac{B^{*}}{B_T}b\right) + 
c_2\sin\left(\frac{B^{*}}{B_T}b\right)
\label{e3-6}
\end{eqnarray}
substituting the above equation into (\ref{e3-5}) it is obtained that
\begin{eqnarray}
\left. \frac{d}{db}\Delta B b^n\right|_{b=b_f} =
-c_1\sin\left(\frac{B^{*}}{B_T}b_f\right) +
c_2\cos\left(\frac{B^{*}}{B_T}b_f\right) = 0,
\label{e3-7}
\end{eqnarray}
Uniqueness requires appropiate boundary conditions. Physically, we expect
that $\Delta B$ reaches its maximum value $D$ at the flux front

\begin{eqnarray}
\left. \Delta Bb^n\right|_{b=b_f} = Db_f^n = 
c_1\cos\left(\frac{B^{*}}{B_T}b_f\right) + 
c_2\sin\left(\frac{B^{*}}{B_T}b_f\right).
\label{e3-8}
\end{eqnarray}

this condition, together with the former result, corresponds to a Cauchy boundary condition.
With (\ref{e3-7}) and (\ref{e3-8}) we obtain  $c_1=Db_f^n\cos(B^*/B_Tb_f)$ and
$c_2=Db_f^n\sin(B^*/B_Tb_f)$, substituing both into (\ref{e3-6}) we find the following solution for $\Delta B$:
\begin{eqnarray}
\Delta B= D\frac{\left( 1+\frac{B_f}{B^*} \right)^n}{\left( 1+\frac{B}{B^*}
\right)^n}
\cos \left( \frac{B-B_f}{B_T} \right).
\label{e3-9}
\end{eqnarray}

Since $\Delta B = 0$ is fullfilled at the sample boundaries,  the applied field must satisfy that

\begin{eqnarray}
\mu_0H_{fi,m} = \mu_0H_a = B_T(2m+1)\frac{\pi}{2} + B(x_f);
\label{e3-10}
\end{eqnarray}

on the contrary, for every $H_a \neq H_{fi,m}$ one has necessarily that $D=0$. 
 
The field $\mu_0H_{fi,m}$ is known as the instability field which determines the threshold where both Lorentz and pinning forces can support a deviation from 
their equilibrium value.

Expressions for the instability field $\mu_0H_{fi,m}$ and deviation $\Delta B$ can be written for
partial and full penetration states, for that, it is important to know first the flux front position $x_f$ for each state. For partial penetrate (PP) states, $x_f$ is given by (\ref{e2-8}) fulfilling that $B(x_f)=0$; for full penetrate (FP) states $x_f=d/2$ and $B(d/2)=B_h$, see equation (\ref{e2-9}), thus $\Delta B$ and $\mu_0H_{fi,m}$ can be
rewritten at both zones as follows:

\begin{eqnarray}
\Delta B = \left\{ \begin{array}{lr}
D\displaystyle\frac{1}{\left( 1+\frac{B}{B^*} \right)^n} \cos \left(
\frac{B}{B_T} \right)
&	\mathrm{PP},  \\
D\displaystyle\frac{\left( 1+\frac{B_h}{B^*} \right)^n}{\left( 1+\frac{B}{B^*}
\right)^n}\cos \left( \frac{B-B_h}{B_T} \right)
&	\mathrm{FP},
\end{array}
      \right.
\label{e3-11}
\end{eqnarray}
and 
\begin{eqnarray}
\mu_0H_{fi,m}  = \left\{ \begin{array}{lr}
\mu_0 H_W = B_T(2m+1)\pi/2 
&	\mathrm{PP},  \\
 \mu_0H_W + B_h 
&	\mathrm{FP}.
\end{array}
      \right.
\label{e3-12}
\end{eqnarray}
At the FP states, the magnetic induction at the center of the sample is
given by
\begin{eqnarray}
\frac{B_h}{B^*}  = 
\left(\left( 1 + \frac{\mu_0H_{fi,m}}{B^*} \right)^{n+1}\!\!\!\!\!\!-
\alpha(n+1)\frac{B_0}{B^*}
\right)^{\frac{1}{n+1}} -1, 
\label{e3-13}
\end{eqnarray}
here $B_0=\mu_0j_0d/2$ is the well-known Bean's penetration field. For
PP states $\mu_0H_{fi,m}$ is dominated by the thermal
characteristics over the magnetic properties, this result agrees to the Wipf's results 
\cite{PhysRev.161.404} for $m=0$. On the other hand, for FP states $\mu_0H_{fi,m}$ is strongly  correlated with the magnetic properties so it is necessary to solve 
the equations (\ref{e3-12}) and (\ref{e3-13}) numerically. 
%
\section{Example: Comparison with experimental data} 
\label{sec:4}
%
In order to validate the theoretical methodology presented in the former sections, here are used the parameters and experimental data obtained by Chabanenko et~al \cite{Chaba2013}; specifically, the required information were extracted from the magnetization hysteresis at $T_B = 5.9\mathrm{K}$. They proposed $\alpha$ as a linear function of the temperature 
\begin{eqnarray*}
\alpha = \displaystyle\frac{T_0-T}{T_0}
\end{eqnarray*} 
with $T_0 =13.2\mathrm{K}$, and a heat capacity with no magnetic induction dependence
\begin{eqnarray*}
C(T) = \alpha_0T^3 + A_0e^{-\delta(T)/T}
\end{eqnarray*}
where $\alpha_0= 3.7\;\mathrm{J/(m^3K^4)}$, $A_0=$ and $\Delta = 38.1\mathrm{K}$. 
Finally, the best Kim-Anderson parameters were $n = 6 $, $B^*=10.1$;  with such values the equation (\ref{e2-10}) provides the penetration field $\mu_0H_P = 1.12$.
\subsection{The $H-T$ instability diagram}
\begin{figure}[ht!]
\centering 
\includegraphics[trim = 5mm 5mm 5mm 5mm, clip,width=0.75\textwidth]{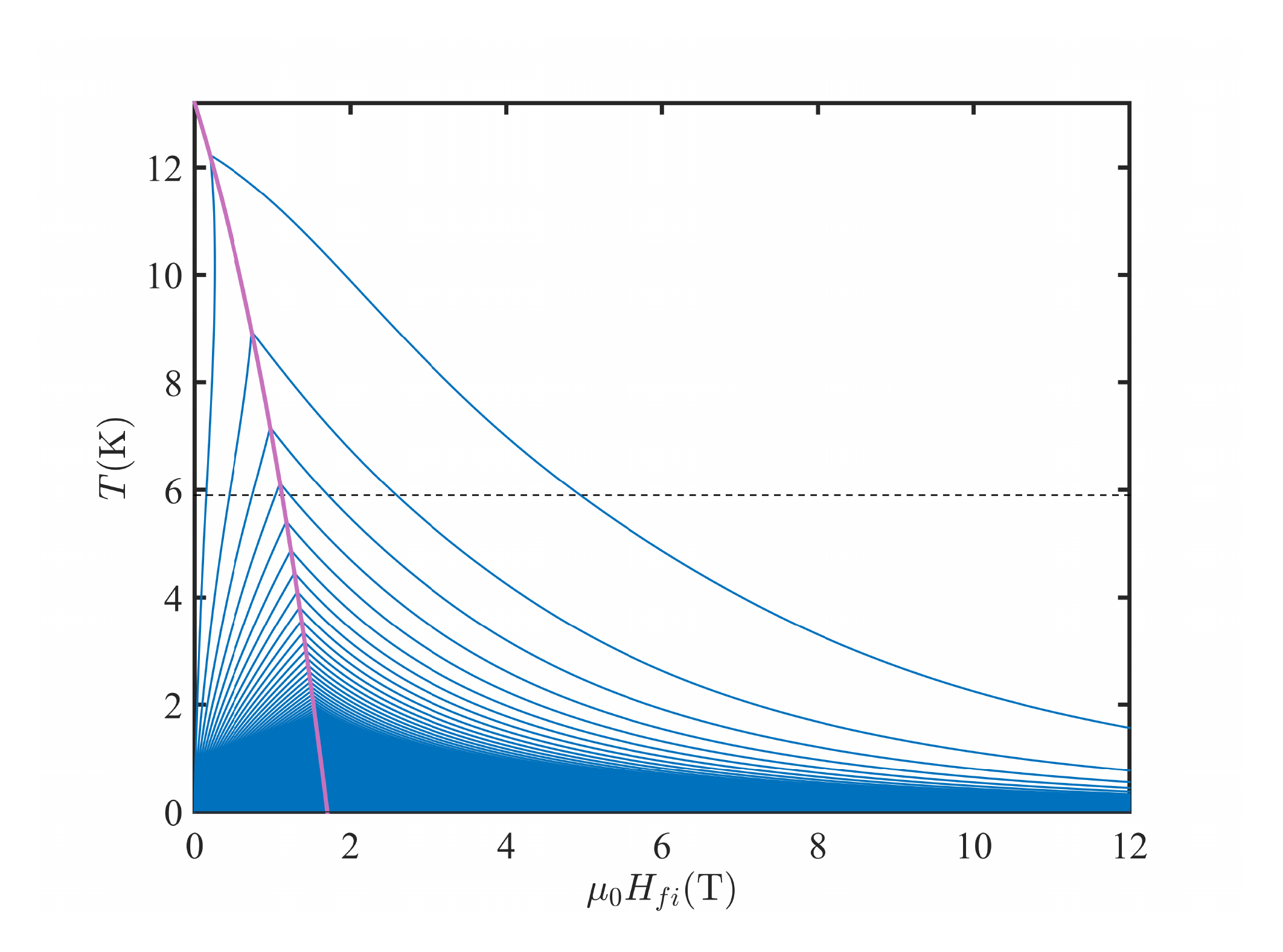}
\caption{The $H-T$ instability diagram. The  parameters employed to obtain this diagram are presented at the 
beginning of section \ref{sec:4}. The purple line is the first penetration field 
$\mu_0H_P$ for different temperatures. The blue curves represent the  
instability field branches $\mu_0H_{fi,m}$ for $m=0,1,2,\dots$. As can be seen, as the 
temperature decreases, the branches tend to spliced. The dash black line denotes the temperature of the thermal bath $T_B=5.9\mathrm{K}$.
}
\label{F1}
\end{figure}
The $H-T$ instability diagram presented in the figure \ref{F1} is obtained via the equation (\ref{e3-10}) where the blue curves are the instability field branches $\mu_0H_{fi,m}$.
Notice how the branches tend to spliced as the temperature decreases.  
This fact can related with the superconducting material behavior, it seems that at low temperatures it is unstable for any applied field value, consequently, the avalanches will always happen. 

The PP and FP states are separated by the purple line which corresponds to the first penetration field $\mu_0 H_P$, calculated with the equation (\ref{e2-10}), for different temperatures $T=T_P$. Each branch has a temperature range $(0,T_p]$, the branch $m = 0$ for the PP state 
corresponds to the Wipf's field, it is extended close to the superconducting transition
temperature, and is the only one that presents a change of curvature.

The Wipf's field is also defined for FP states, it is characterized by a change of curvature for 
temperatures close to $ T_P $, this change can be related with the dominant influence of the 
thermal properties respect to the magnetic ones, due to its proximity to the 
superconducting transition temperature. For the rest of the branches, the magnetic properties will dominate. Even though the richness of this instability diagram, it is uncertain the exact applied field value where a flux jump occurs. 

We retrieved from Chabanenko's experiment, the applied magnetic
field rate of $1\mathrm{T/min} \approx 0.02\mathrm{T/s}$, even so the 
step size was not reported, here it was estimated as $0.02\mathrm{T}$ with 
 an error of $\pm0.005$. 

The dashed black line denotes the temperature of the thermal bath $T_B=5.9\mathrm{K}$.
The $H-T$ diagram shows that for PP states the branches $m=0,1,2,3$ correspond to
$\mu_0H_P >\mu_0H_{fi,m}=0.17; 0.42; 0.76; 1.01\mathrm{T}$ respectively.
For FP states the equation (\ref{e3-12}) has real solutions only for the branches 
$m=3,2,1,0$ which are $\mu_0H_ {fi,m}=1.23;1.71;2.59;4.95\;\mathrm{T}$, respectively. 
The table \ref{table1} contains the above results as well as the experimental data of $\mu_0H_ {fi,m}$. 

The experimental magnetization curve presents three instability fields
$\mu_0H_ {fi}=0.45,1.09,2.61\mathrm{T}$ at the first quadrant. As can be appreciated, 
the fields $\mu_0H_{fi,1}=0.42\mathrm{T}$, $\mu_0H_{f,3}=1.01\mathrm{T}$ for 
states PP and $\mu_0H_ { fi,1}=2.59\mathrm{T}$ for FP states are in good agreement 
with the experimental data.

As usual, the third quadrant of the curve is pretty similar to the first quadrant. In the third quadrant there are  four instability fields 
$|\mu_0H_{fi}|=0.15,0.44,0.75,1.05\;\mathrm{T}$, they were estimated using a graphic digitalization procedure. The third instability field is an abrupt slope change of the curve rather than a conventional flux jump. The experimental and theoretical values almost match, such closeness suggests that indeed occurred an instability, however the flux jump was frustrated by some unknown physical mechanism.
\begin{table}[ht!]
\begin{center}
\begin{tabular}{|lcc|cc|}
\hline
\multicolumn{5}{c}{Instability field $|\mu_0H_{fi}|\;\mathrm{(T)}$ at $5.9\mathrm{K}$ } \\
\hline
\multicolumn{3}{|c|}{Theory($\mu_0H_{fi,m}$)}      	&\multicolumn{2}{c|}{Experiment\cite{Chaba2013}} 	 	\\
 \hline
\multicolumn{3}{|c|}{Quadrant 1}     		&Quadrant 1 & Quadrant 3 			    			\\
 m	& 	PP 	 &FP	     		&			& 		    			\\
 0 	& 0.17 &4.95				& 	0.45 	& 0.15 						    \\
 1 	& 0.42 &2.59 				& 	1.09	& 0.44 							\\
 2 	& 0.76	&1.71				& 	2.61	& 0.75$^*$ 				 		\\
 3 	& 1.01 &1.22 				&	-		& 1.05 							\\
\hline
\multicolumn{5}{l}{\parbox[t]{8cm}{PP: partial penetrate. FP: Full penetrate.\\
($^*$) This instability field is estimated where the magnetization curve has 
an abrupt change.}}
\end{tabular}
\end{center}
\caption{Theoretical and experimental values of the instability field $\mu_0H_{fi,m}$}
\label{table1}
\end{table}
%
%
\subsection{Magnetic induction profiles}
Once the external instability fields have been found, it is interesting to know the magnetic induction behavior.  Figure \ref{F2} shows the magnetic induction of the superconducting material precisely at the instability field, also the isothermal as well as the disturbed fields are graphed. Panel (a): The black lines are the isothermal field 
distribution $ B $, while the blue and red lines are the disturbed field corresponding to PP and FP states, respectively. Panels (b) and (c) show the $\Delta B$ oscillatory 
behavior, for PP and FP states respectively; to magnify $\Delta B$ it was used 
a perturbation amplitude $D=0.1(\mathrm{T})$. 

The perturbed magnetic induction resembles a flexible cantilever structure 
with its fixed edge at the superconductor boundary, and maximum amplitude 
$D$. For PP states the oscillations increase as the instability field increases 
because the external field suffers a great opposition to diffuse into 
the superconducting material, unlike for FP states where occurs the opposite.

Although the graphs shape suggests that instability fields resonances may occur,
it is important to point out that our theory exclude them, however the oscillations are only present at the instabilities. For another external field 
value, the magnetic induction will not suffer any deviation from its isothermal 
behavior. 

The oscillations promote or inhibit the magnetization, this fact can be appreciated in panel (a).
 There are regions where the incursion of $B$ is encouraged and
other ones where is mitigated, at this scenario, the current density magnitude
can approach to zero once the temperature reaches its maximum value, while if the
temperature decreases, the current density value is increased.
%
\subsection{Temperature change}
%
%
\begin{figure}[ht!]
\centering 
\includegraphics[trim = 10mm 10mm 10mm 10mm, clip,width=0.85\textwidth]{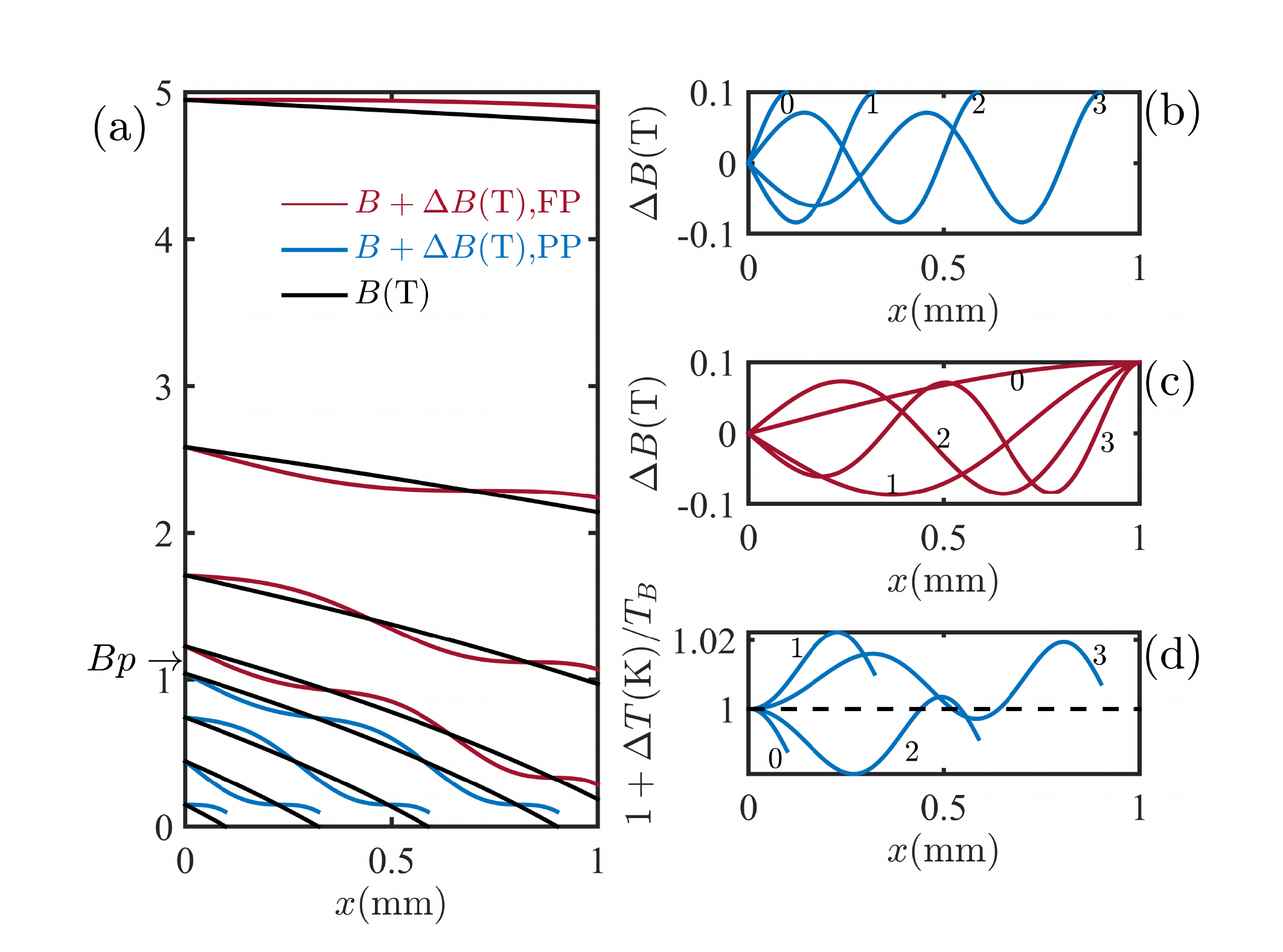}
\caption{For a temperature $T_B = 5.9\mathrm{K}$ and a penetration field $\mu_0H_P=1.12\mathrm{T}$
there are eight instability fields given by $\mu_0H_{fi,m}=B_T(2m+1)\pi/2+B_H$. For states PP 
$m=0,1,2,3$, then $\mu_0H_{fi,m}=0.17; 0.42; 0.76; 1.01\mathrm{T}$, and for states FP, $m=3,2,1,0$
then $\mu_0H_{fi,m} = 1.23; 1.71; 2.59; 4.95\;\mathrm{T}$ (see table \ref{table1}).
Panel (a): the magnetic induction profiles $ B + ΔB $ are shown, for PP they are the blue lines and for FP
they are the red lines, while the black lines correspond only to $ B $. Panels (b) and (c) show the behavior
of the deviation $ ΔB $ inside the sample for PP and FP states respectively. 
We used a perturbation amplitude $D=0.1(\mathrm{T})$ to magnify $\Delta B$.
Panel (d) shows the change of temperature around the temperature of the thermal bath for PP states, using $D=0.0001(\mathrm{T})$ is obtained a maximum deviation of 2\%.
}
\label{F2}
\end{figure}

This example finished with the description of the temperature. The theory describes a temperature change $\Delta T$ produced by Joule's heat, in the adiabatic regime. This causes that the reference isothermal state becomes a non-equilibrium disturbed state. At this regime the temperature still be not homogeneous,  this fact is illustrated by calculating 
$\Delta T$. Given the boundary condition $T(x=0)=T_B$, the local temperature is $T = T_B - \Delta T +\Delta T(x=0)$, and using the equations (\ref {e3-3}) and (\ref {e3-9}) for PP states, it is found the following:
\begin{eqnarray*}
 T = T_B + \frac{\alpha}{\partial_T\alpha}\frac{D}{B_T}
\left(1+\displaystyle\frac{B_f}{B^*}\right)^{\displaystyle\!\!n}\!\!
\left( 
\frac{\mathrm{sgn}\left(\displaystyle\sin{\frac{B-B_f}{B_T}}\right)}{\left(1+\displaystyle\frac{\mu_0H_{if,m}}{B^*}\right)^{\displaystyle\!\!n}\!\!} -
\frac{\displaystyle\sin{\frac{B-B_f}{B_T}}}{\left(1+\displaystyle\frac{B}{B^*}\right)^{\displaystyle\!\!n}\!\!}
\right). 
\end{eqnarray*}
As is expected, the theory predicts for the temperature an oscillating flexible cantilever behavior, in the adiabatic regime. Panel (d) of figure \ref{F2} shows the temperature profile around the   thermal bath temperature for PP states; employing the value $D=0.0001(\mathrm{T})$ is obtained a maximum deviation of 2\%. The maximum temperature increment is determined by the intensity of the perturbation $ D $, in our theoretical calculation we assume that it is a constant quantity but  it must depend on the external field and the thermal properties of material.
%
\section{Conclusion}
\label{sec:5}
%
Starting from a simple but strong physical scheme the
thermomagnetic stability of type-II superconductors is studied, this allows to find the instability field $H_{fi,m}$ for external magnetic fields applied with arbitrary direction at
the $yz$ sample plane.

With this theory is possible to construct $H-T$ maps for both PP and FP states to accurately predict the instability field $H_{fi,m}$ according to the thermal bath and field applied values, this result marks differences from Wipf's scheme \cite{PhysRev.161.404} since his approach was constrained to find the only one solution $H_{fi,0}$ for PP states, meanwhile in this work one can obtain a rich branches set for $H_{fi,m}$, going further up to FP states.

Since the theorical study is a first order approximation, the deviation behavior is periodic along the $ x $ direction. It is constructed critical state magnetic induction profiles  that resemble a  flexible cantilever structures. 

The instability field indicates a flux jump occurrence, and the theory presented here found that an ondulatory profile corresponds to a flux jump. Even so is not possible during an experiment to determine accurately the instability field value where a flux jump appears, if the step size $\Delta H$ is changed, there is a neighborhood around the theoretical value where can be detected a jump during the experiment. Thus it is reasonable to think that the oscillatory behavior can be replicated to profiles around the instability  field. 

There remains the challenging to consider an anisotropic critical state model\cite{0953-2048-26-12-125001,0953-2048-29-4-045004}, both thermal and conductive properties of the heat capacity $C(T,H)$, as well as other sample geometry\cite{xia2017numerical}.
%
\begin{acknowledgements}
The authors acknowledge Dr. Felipe P\'erez-Rodr\'iguez invaluable feedback
during the development of this paper. We thank M.C. Noe T. Tapia Bonilla for his careful comments.
\end{acknowledgements}
%
\appendix
\section{Temperature change $\Delta T$}
\label{sec:6}
%
The temperature change $\Delta T$ is obtained from the heat equation 
\begin{eqnarray*}
C(T)\frac{dT}{dt} = \nabla\cdot(\alpha(T)\nabla T) + \frac{\partial Q}{\partial t},
\end{eqnarray*}
integrating from time and assuming the adiabatic regime, where
$\nabla\cdot(\alpha(T)\nabla T)\approx 0$ and $\partial_tC(T)\approx 0$, it is
obtained that
\begin{eqnarray}
C(T)\Delta T = \Delta Q, 
\label{ec-A1}
\end{eqnarray}
here $\Delta Q$ corresponds to the heat change generated by the critical current
density and the electromotive force produced by the magnetic induction profiles
changes. To obtain an analytic expression for the heat change, we use the
Faraday induction law as follows:
\begin{eqnarray*}
E_z =\int \frac{\partial B_y}{\partial t}dx + E_{z0}, \quad
E_y =-\int \frac{\partial B_z}{\partial t}dx + E_{y0}. \nonumber 
\end{eqnarray*}
Substituting the former equations into Joule's heat $dQ/dT =
\mathbf{E}\cdot\mathbf{j}$, one has that
\begin{eqnarray*}
\frac{dQ}{dt} = -j_y\!\!\int\!\!\frac{\partial B_z}{\partial t}dx +
 					  j_z\!\!\int\!\!\frac{\partial B_y}{\partial t}dx 	+
					  j_yE_{y0} + j_zE_{z0}	\nonumber 
\end{eqnarray*}
Employing the relations 
\begin{eqnarray}
j_y = j_c\cos{\varphi} & &  j_z = j_c\sin{\varphi} 	\nonumber \\
B_y= B\cos{\phi} 	   & & B_z = B\sin{\phi}		\nonumber 
\end{eqnarray}
we have that
\begin{eqnarray*}
\frac{dQ}{dt}  = j_c\sin{(\varphi -\phi)}\int \frac{\partial B}{\partial t}dx +
j_c(E_{y0}\cos\varphi + E_{z0}\sin\varphi) \nonumber
\end{eqnarray*}
The last term can be associated to the background heat produced by a constant
voltage applied to the superconductor. In our study we assume that
$E_{y0}=E_{z0}=0$, besides, the material is isotropic thus $\varphi = \phi +
\pi/2$. With the above information, the heat change takes the final form:
\begin{eqnarray}
\Delta Q & = 		& 	\int j_c\left(\int\frac{\partial B}{\partial t}dx\right)dt
		    = j_c\int\!\! \Delta Bdx + 
		       \int\left(\int\frac{\partial B}{\partial t}dx\right)\frac{\partial j_c}{\partial t}dt, \nonumber \\
		   &\approx	& 	j_c\int\!\! \Delta Bdx, \nonumber
\end{eqnarray}
where $\partial_t j_c\approx 0$. Finally, substituting the last equation into (\ref{ec-A1}) 
we obtain:
\begin{eqnarray}
\Delta T C(T) = j_c\int_{x}^{x_0}\!\! \Delta Bdx. \nonumber
\end{eqnarray}
%
\section*{References}
\bibliography{OAHF_bibfile}{}
\bibliographystyle{plain}
%
\end{document}